\documentstyle[11pt,paspconf,epsf,psfig,twoside]{article}

\begin{document}

\def\Mdot{\hbox{$\dot M$}}
\def\Msun{\hbox{$M_\odot$\,}}
\def\Rwd{\hbox{$R_{_{\rm WD}}$\,}}

\title{Continuum and Line emission from Post-shock Flows}

\author{Mark Cropper$^{1}$, Kinwah Wu$^{2}$ \& Gavin Ramsay$^{1}$}
\affil{
$^{1}$ Mullard Space Science Laboratory, University College London,
             Holmbury St Mary, Dorking, RH5 6NT, UK \\
$^{2}$ Research Centre for Theoretical Astrophysics, School of Physics,
             Sydney University, Sydney NSW, Australia.}

\begin{abstract}
The post-shock flow in the accretion region of MCVs is considered to be a
magnetically confined cooling flow. The X-ray spectrum from this hot plasma is
emitted from a range of temperatures and densities. Extracting information from
such spectra obtained with the next generation of X-ray satellites will rely on
models for the structure of this flow which include appropriate treatment of
the cooling mechanisms, boundary conditions and of absorptions and reflection
components. This paper summarises the current situation and explores some of
the difficulties arising in the exploitation of the line emission from these
systems. It concludes by looking at improvements to the models now in the
pipeline.
\end{abstract}

\section{Introduction}

Since the earliest availability of X-ray data on AM Her systems, efforts
have been made to characterise their X-ray emission. Fits have generally been
made using a single temperature bremsstrahlung component (where one is
detected) together with a softer reprocessed component from the surface of the
white dwarf (Fabian, Pringle \& Rees 1976). In terms of their model of a 1-d
magnetically collimated flow, the harder (bremsstrahlung) component contains
information on the hot gas in the region between the shock front at which the
flow becomes subsonic and the surface of the white dwarf. Although a
prescription of the temperature and density profile in the gas between these
boundaries has been available for some time (Aizu 1973), and several authors
have constructed spectra using the Aizu and more detailed descriptions (for
example Imamura et\,al. 1987, van Teeseling et\,al., these proceedings) it is
only recently that this knowledge has been used to extract information from the
fits to the X-ray data (Done, Osborne \& Beardmore 1995, Cropper, Ramsay \& Wu
1998). The impetus has been the availability of approximate models (such as
appropriate for cooling flows in clusters of galaxies) and the prescription of
Wu (1994) which modifies the Aizu calculations to include the effects of
cyclotron cooling in a tractable form.

The system parameters which determine the temperature and density structure of
the postshock flow (and therefore its spectrum) are mainly the strength of the
gravitational potential (the mass and radius of the white dwarf), the accretion
rate per unit area and the ratio of the cyclotron to bremsstrahlung cooling
times within the shock (affected by the magnetic field strength). Fits to
X-ray spectra therefore provide fundamental and relatively directly accessible
information about the accretion flow and the white dwarf (Cropper et\,al. 1998;
Tennant et\,al. 1998; Fujimoto \& Ishida 1996). These parameters are at the
basis of any understanding of the system and of the accretion physics, and it
is clear, especially as the quality of the X-ray data improves with the next
generation of observatories, that this line of attack holds substantial
promise.

A number of assumptions are generally made in the extraction of this
information, which have different consequences depending on the nature of the
system and the particular aspect of the problem under study. These assumptions
need to be relaxed or made more realistic. A common first-order
misconception, for example, is to attribute the characteristic temperature of
the single temperature bremsstrahlung fit to the shock temperature. Since most
of the X-ray emission is from regions near the base of the flow where the
density is highest, this temperature will be less than half that of the
immediate post-shock flow, and white dwarf masses will be significantly
underestimated. It is possible to do considerably better, for example by
applying the calculations of Wu (1994) as in Cropper et al. (1998). Even so,
there are indications (Ramsay et\,al. 1998) that these overestimate the white
dwarf mass. It is possible also to use the line emission from the hot optically
thin flow (Ishida, these procedings), but these also have their difficulties
(see below and van Teeseling et\,al., these proceedings). The development of
the tools for the extraction of this information is, therefore, at the present
stage, an ongoing activity.

\section{Continuum fits}

The basis for this technique is that streaming velocities of the infalling
material are randomised by the shock, so that bulk motions, created by the
transfer from gravitational potential energy to kinetic energy, are converted
to thermal motions. Applying the strong shock conditions (where the velocity
drops by a factor of 4 across the shock) it is easy to show that
$T_{s}\sim3GM\mu m_{H}/8kR$ where $T_{s}$ is the shock temperature, $G$ is the
gravitational constant, $M$ and $R$ are the mass and radius of the white dwarf,
$\mu$ is the mean molecular mass, $m_{H}$ is the mass of a Hydrogen atom and
$k$ is Boltzmann's constant.

The equations used for setting the temperature and density structure 
are 1-d stationary state hydrodynamic conservation equations of mass, 
momentum and energy (see Wu 1994). In the formulation, the cyclotron 
cooling is included in a power-law form. It is derived from the 
assumption that the emission escapes from the sides of the accretion 
column, as appropriate for the cyclotron beaming, and the cyclotron 
cooling is contributed only by that emission which is optically 
thick, up to a peak frequency that depends on the local temperature and 
the electron number density (see Saxton, Wu \& Pongracic 1997). 
The shock transition is adiabatic, and the pre-shock flow is highly 
supersonic, such that the strong shock condition is applicable. The 
electrons and the ions are strongly coupled to ensure an equal 
temperature for both the electrons and ions. Two-fluid treatments
exist in the literature (for example Imamura et\,al.\ 1987; Wu \& Saxton, 
these proceedings; van Teeseling et\,al., this proceedings): there 
the issues of exchange timescales between the electrons and ions have 
to be treated carefully (see Wu, in preparation).  

Figure 1a shows the temperature and density profile for the bremsstrahlung only
case (Aizu profile) for three white dwarf masses and a pure Hydrogen plasma
($\mu=0.5$). It is instructive to note that the shock temperature of a 0.6
\Msun white dwarf the shock temperature is 17 keV whereas the equivalent for
1.0 \Msun is substantially higher at 47 keV. Moreover, the shock height is more
than 4 times higher: 0.38 \Rwd. 

\begin{figure}
\hspace*{-5mm}
\begin{minipage}{72.5mm}
\psfig{file=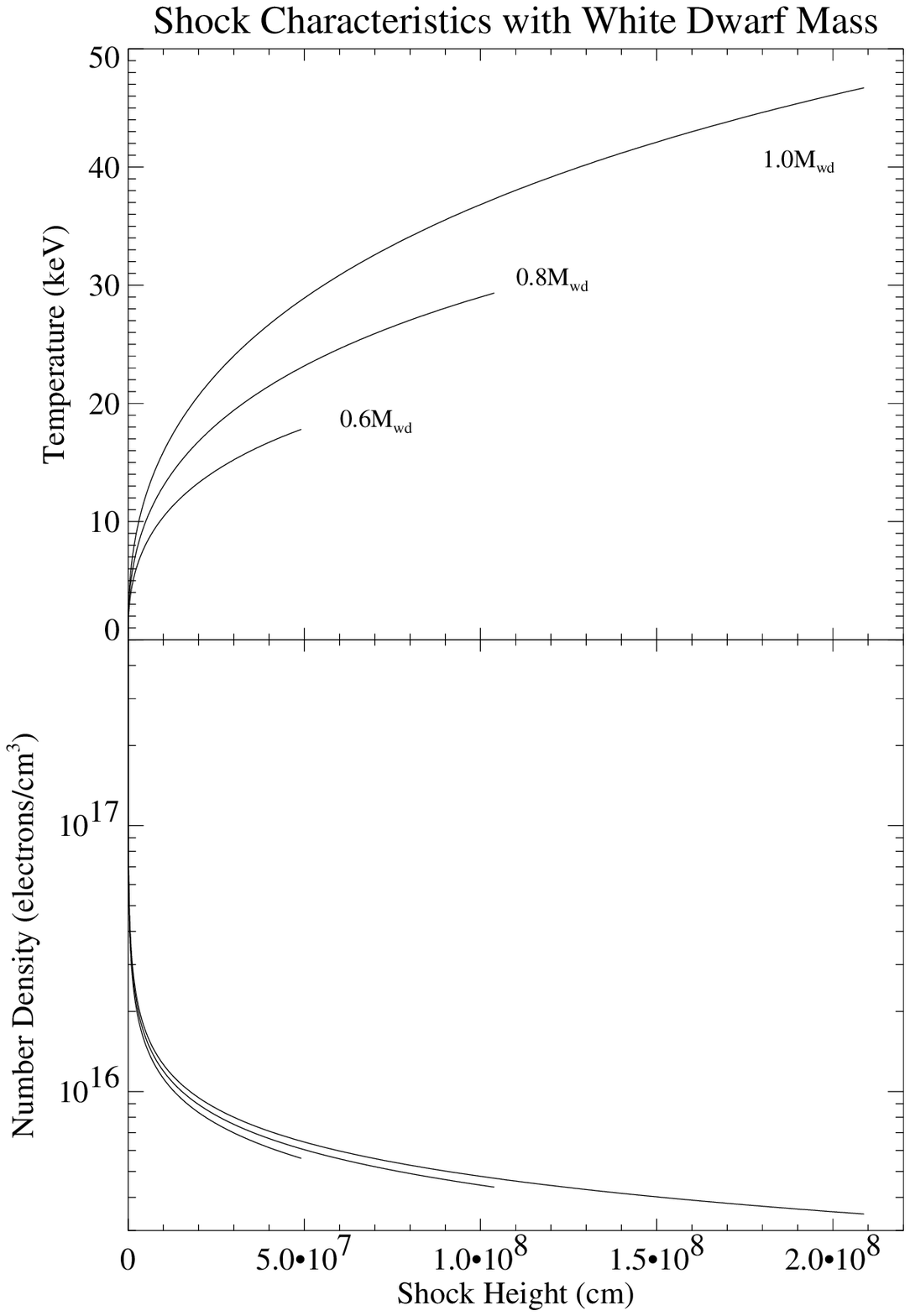,width=72.5mm}
\end{minipage}
\hspace*{-1cm}
\begin{minipage}{72.5mm}
\psfig{file=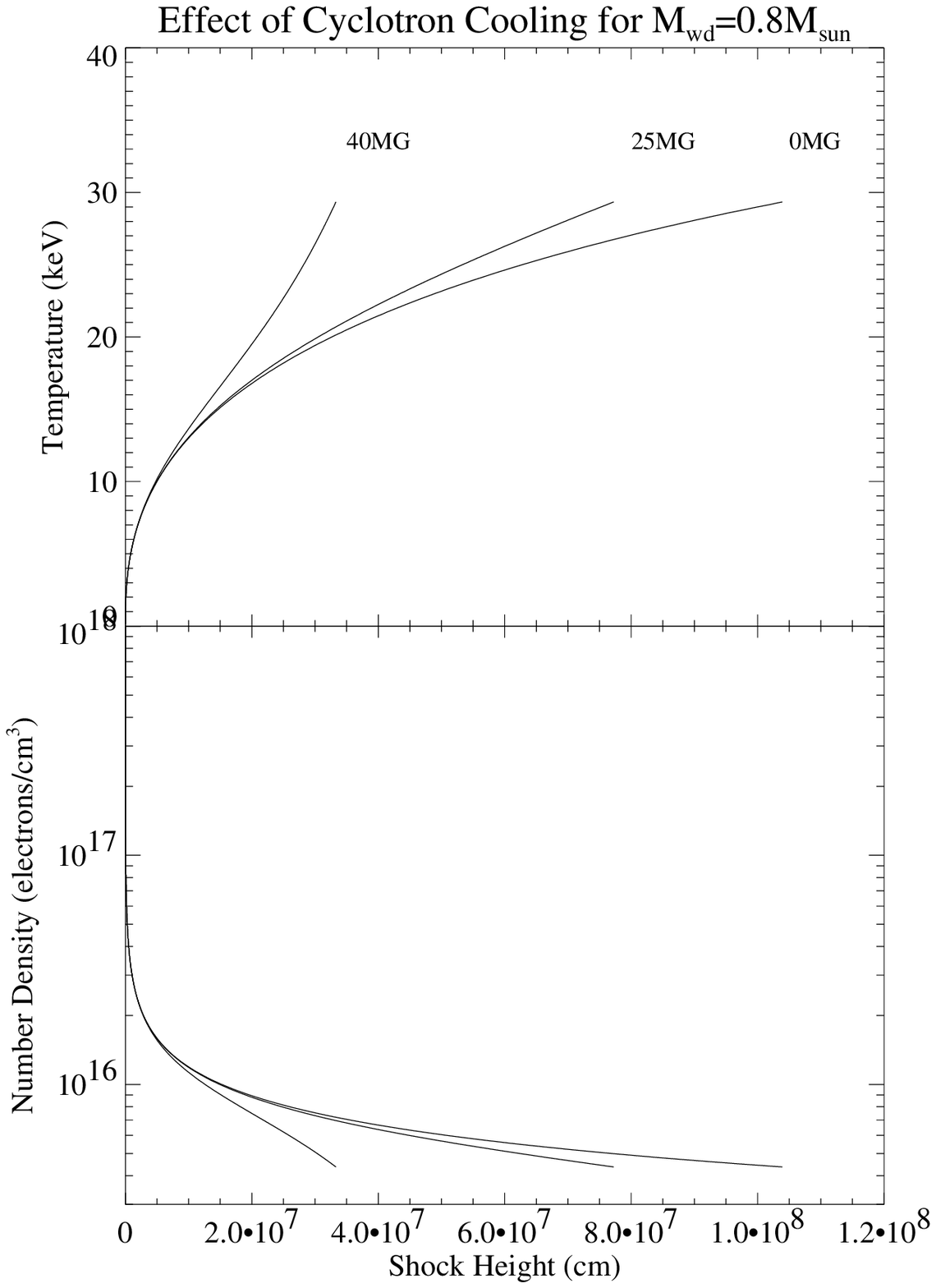,width=72.5mm}
\end{minipage}
\caption{(a) The left hand figure shows the temperature and density profiles
within the postshock flow for different mass white dwarfs accreting at 1
g/s/cm$^2$ for a pure Hydrogen plasma. (b) The right hand figure shows the
dependence of the shock height and temperature and density profiles as a
function of magnetic field, for the same accretion rate and for the 0.8 \Msun
case.}
\end{figure}

Figure 1b shows the effect of an increasing magnetic field on the shock
structure. The initial shock temperature is unchanged (because it reflects
simply the thermalised potential energy), but the height of the shock is
significantly decreased as the additional cooling mechanism removes energy in
the hottest part of the postshock flow. Thus polars will have a shock height
typically only half of that of IPs for the same local mass accretion rates and
white dwarf masses.

With the knowledge of the temperature and density structure as a function of
height, it is possible to construct the overall emission spectrum from the
flow. It is easy to show that except for the very base of the flow, where in
any case the complicating effects of the white dwarf photosphere begin to be
important, the emission is optically thin in the continuum and most lines, so
that the resultant spectrum is simply the sum of the local spectra emitted as a
function of height in the postshock flow. Cropper et\,al. (1998) used the
revised MEKAL models, which are optically thin opacity models, available in
XSPEC. The resulting spectrum is shown in Figure 2.

\begin{figure}
\hspace*{-5mm}
\begin{minipage}{72.5mm}
\psfig{file=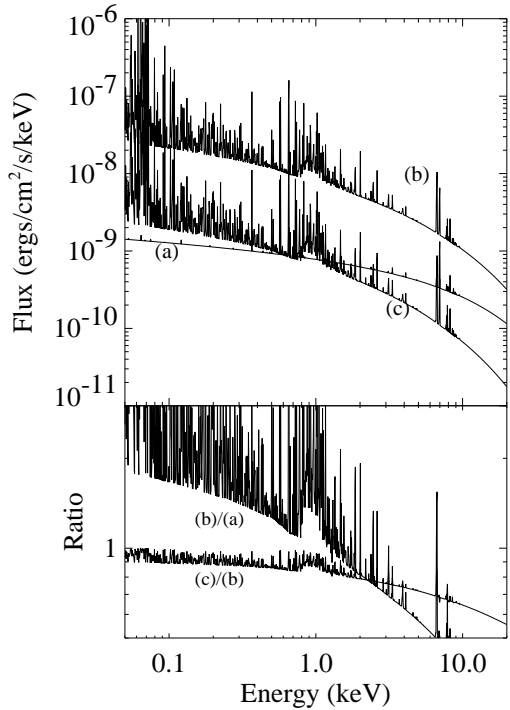,width=72.5mm}
\caption{Spectra from optically thin models of the postshock flow for a 0.6
\Msun white dwarf accreting at 1 g/s/cm$^2$. (a) is the single temperature
spectrum from the immediate postshock region; (b) is the multi-temperature
spectrum (displaced vertically by a factor 10 for clarity) but with no magnetic
field and (c) is the same but including a 40 MG field.}
\end{minipage}
\begin{minipage}{62.5mm}
\scriptsize
\begin{tabular}{lrrr}
Source    &   CRW98   &    revised   &    difference \\ 
          & $\mu=0.5$ & $\mu=0.615$  &               \\ \tableline
          &           &              &               \\
AM Her    &      1.22 &       0.94   &          0.28 \\
EF Eri    &      0.88 &       0.78   &          0.10 \\
BY Cam    &      1.08 &       1.27   &        --0.19 \\
V834 Cen  &      0.51 &       0.51   &          0.00 \\
QQ Vul    &      1.12 &       1.30   &        --0.18 \\
          &           &              &               \\
EX Hya    &      0.52 &       0.45   &          0.07 \\
AO Psc    &      0.40 &       0.45   &        --0.05 \\
FO Aqr    &      1.22 &       1.11   &          0.11 \\
TV Col    &      1.20 &       1.21   &        --0.01 \\
BG CMi    &      1.25 &       1.36   &        --0.11 \\
TX Col    &      0.55 &       0.48   &          0.07 \\
PQ Gem    &      1.35 &       1.32   &          0.03 \\
AE Aqr    &      0.30 &       0.62   &        --0,32 \\ 
          &           &              &               \\ \tableline
          &           &              &               \\
          &           &              &               \\
          &           &              &               \\
\end{tabular}\\
\normalsize
Table 1: Revised masses in \Msun for the systems observed with {\it Ginga} in
Cropper et\,al. (1998).
\end{minipage}
\end{figure}

Figure 2 indicates the importance of including the effect of the temperature
and density gradients in the postshock flow: the multi-temperature spectrum is
significantly softer. The lower plot shows that the addition of the cyclotron
emission further softens the spectrum. White dwarf masses derived using (c)
will therefore be higher than those from (b) and significantly higher than
those from (a).

Figure 2, taken from Cropper et\,al. (1998) is internally inconsistent, in that
it assumes a pure hydrogen plasma to set the temperature and density structure,
but uses plasmas of cosmic composition $\mu=0.615$ to calculate the emission
spectra. Using a cosmic plasma for the flow increases the shock temperature by
a factor 1.23, significantly hardening the spectrum. This initially works to
reduce derived white dwarf masses. Refitting the {\it Ginga} data in Cropper
et\,al. (1998) results in revised masses given in Table 1. The 90\% confidence
ranges for the masses range from 0.15 \Msun in the best cases (eg AM Her) to
0.75 in the worst (AE Aqr, V834 Cen). The masses are not uniformly reduced
because of the interaction between the emitted spectrum and the absorbers. The
major changes are the decrease in mass for AM Her, in which XSPEC has found a
different minimum in the $\chi^{2}$ plane, and an increase in the mass for AE
Aqr (which in any case was not previously well constrained).

The fits in Table 1 predict the {\it Ginga} spectra well and strong constraints
on the masses can be found for those systems with sufficient data quality. The
model includes, in addition to the multi-temperature emission, the reflection
from the white dwarf surface and the effect of a warm (ionised)
absorber. Nevertheless, as noted in the introduction, there are additional
effects to be included which will modify the derived masses, probably at the
0.1 -- 0.2 \Msun level. These include instrumental calibration, more complex
absorptions including absorption by dense filaments (Done \& Magdziarz 1998),
and improvements to the model assumptions (see Section 4 below). It remains to
be seen whether these effects work to reduce the derived masses, some of
which are at the top end of expectations.

\section{Line fits}

The temperature and density structure calculated for the continuum fits can
also be used to determine the line emissivity and ionisation profiles as a
function of height within the postshock flow (Ishida, these proceedings, Wu,
Cropper \& Ramsay, in prep). These calculations indicate that even for those
species with high ionisation energies, the line emission is from the lowest
part of the postshock flow, near the white dwarf. For a 1 \Msun white dwarf and
no magnetic field, the Fe\,{\sc xxvi} Ly-$\alpha$ emissivity peaks below 1\% of
the height of the column; for a typical field found in polars, the peak is at
$\sim 1-2$\%. For lower mass white dwarfs the situation is better: the
equivalent values for a 0.5 \Msun white dwarf is $\sim10$\% and $\sim30$\%.

Concerns on the physics of the optically thin models aside (van Teeseling
et\,al., these proceedings), this indicates that measurements of the emission
lines in the X-ray spectra can provide constraints on mass derivations most
readily for low mass white dwarfs. Comparison of the masses derived using both
X-ray methods for EX Hya (Table 1 and Ishida, these proceedings) finds close
agreement at $\sim0.5$\Msun. It is clear that at higher masses the situation is
more difficult, and may await a more appropriate treatment of the boundary
conditions at the base of the flow. Optical depths, including those from
scattering, need to be considered. In addition, the heated photosphere around
the base of the postshock flow may absorb much of the line emission. Evidence
for this is available from EUVE spectra (Mauche et\,al. 1995) where the
expected forest of emission lines seen in Figure 2 is absent. A further
complicating effect is the fluorescence from the heated white dwarf photosphere
(see van Teeseling, Kaastra \& Heise 1996), especially from iron. The
implications are that, at least for more massive white dwarfs, both excellent
data {\it and} detailed modelling of these complicating effects will be
necessary in order to extract information from the X-ray line spectra.

\section{Improvements -- including the gravitational potential}

One of the main shortfalls of the 1-d calculations is that the gravitational
potential is not included implicitly in the momentum and energy equations, but
enters only as a boundary condition -- the assumption that the pre-shock flow
velocity takes the value of the free-fall velocity at the surface of the white
dwarf. This is adequate for low mass white dwarfs with high local accretion
rates, or where the magnetic field is strong, but Figure 1 indicates that the
shock height is indeed significant, especially for masses above 1 \Msun. The
freefall velocity at this point is significantly less than that at the surface,
so that the shock temperature will be lower. However, gravitational potential
energy is still available for release in the tall postshock region, so that the
subsequent temperature profile can be expected to exceed that of the Aizu
profile.

Including the gravitational potential complicates the solution of the
hydrodynamic equations, but a stepwise scheme can be constructed consisting of
a pair of first order nonlinear ODEs to determine velocity and a pressure-like
term as a function of height (Cropper et\,al., in prep). Unfortunately it is a
boundary value rather than initial value problem, and the upper boundary value
is itself a product of the calculation, but nevertheless the behaviour is well
behaved and the calculations converge after only a few iterations. 

\renewcommand{\floatpagefraction}{0.99}
\begin{figure}
\psfig{file=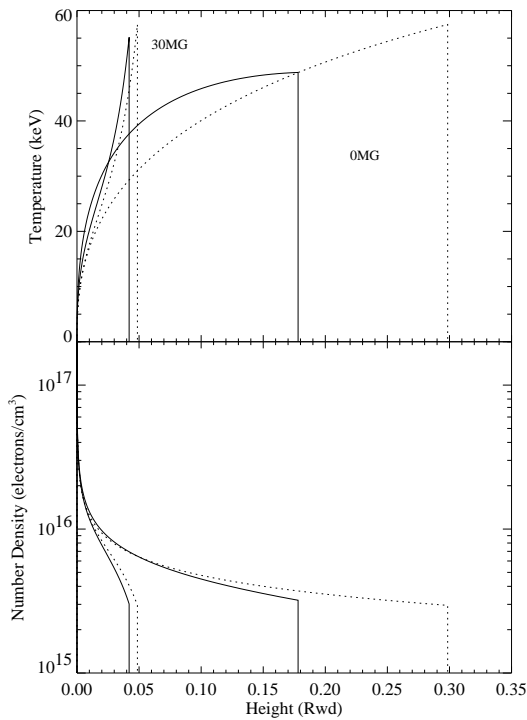,height=110mm}
\caption{Upper plot: The temperature profile for the case including the
gravitational potential (solid line) compared to the standard zero gravity
profile (dotted line) when cyclotron cooling is included (30 MG) and negligible
(0MG). The vertical lines are at the position of the shock front. Lower plot:
as above but for the electron number density profiles. See text for details.}
\end{figure}

The results are shown in Figure 3. This shows the characteristics noted above,
resulting in a flatter temperature profile over much of the height of the
postshock flow. As expected, the corrections are less important for those flows
where the cyclotron cooling significantly reduces their height. A consequence
of Figure 3 is that there is more emission from this flow in the intermediate
energy range to 20 keV compared with that from an Aizu profile: this means,
somewhat counterintuitively, that they produce harder spectra, which will act
to reduce the masses derived under the assumptions of Cropper et\,al. (1998)
and Ramsay et\,al. (1998). It should be noted, however, that this is a
generally less significant correction than the inclusion of the cyclotron
cooling, where that is required. The effect is most important for IPs with
primary masses exceeding 1 \Msun (as do several in Table 1). Fits using this
revised profile to the XY Ari data in Ramsay et\,al. (1998) result in a
reduction in mass from 1.28 \Msun to 1.19 \Msun using the {\it Ginga} data, or
from 1.03 \Msun down to 0.89 \Msun using {\it RXTE} data. This is to be
compared with the range of $0.78 - 1.03$\Msun from constraints provided by 
the eclipse duration (Ramsay et\,al. 1998).

\section{Conclusions}

Progress is being made in the modelling of the hard X-ray emission from mCVs.
Although further improvements to the models are required, particularly in
respect of the more appropriate treatment of the lower boundary with the white
dwarf, and of the coupling between ion and electron populations in the
immediate post-shock region, such data already provide direct diagnostics of
the characteristics of the system and of the flow.

\acknowledgments 
KW acknowledges the support from the Australian Research
Council through an Australian Research Fellowship.

\end{document}